\documentclass[twocolumn,english,notitlepage,superscriptaddress]{revtex4-2} 
\usepackage[utf8]{inputenc}

\date{\today}

\usepackage{amsmath}
\usepackage{amssymb}
\usepackage{graphicx}
\usepackage{svg}
\usepackage{color}

\usepackage{physics} 

\usepackage{babel}

\begin{document}


\title{Controlling Photon Entanglement with Mechanical Rotation } 

\author{Marion Cromb}
\affiliation{School of Physics and Astronomy, University of Glasgow, Glasgow,
G12 8QQ, UK}

\author{Sara Restuccia}
\affiliation{School of Physics and Astronomy, University of Glasgow, Glasgow,
G12 8QQ, UK}

\author{Graham M. Gibson}
\affiliation{School of Physics and Astronomy, University of Glasgow, Glasgow,
G12 8QQ, UK}

\author{Marko Toroš}
\affiliation{School of Physics and Astronomy, University of Glasgow, Glasgow,
G12 8QQ, UK}

\author{Miles J. Padgett}
\email{miles.padgett@glasgow.ac.uk}
\affiliation{School of Physics and Astronomy, University of Glasgow, Glasgow,
G12 8QQ, UK}

\author{Daniele Faccio}
\email{daniele.faccio@glasgow.ac.uk}

\affiliation{School of Physics and Astronomy, University of Glasgow, Glasgow,
G12 8QQ, UK}

\begin{abstract}

 Understanding quantum mechanics within curved spacetime is a key stepping stone towards understanding the nature of spacetime itself. Whilst various theoretical models have been developed, 
 it is significantly more challenging to carry out actual experiments that probe quantum mechanics in curved spacetime. 
 By adding Sagnac interferometers into the arms of a Hong-Ou-Mandel (HOM) interferometer that is placed on a mechanically rotating platform,  we show that non-inertial motion modifies the symmetry of an entangled biphoton state. 
 As the platform rotation speed is increased, we observe that HOM interference dips transform into HOM interference peaks. This indicates that the photons pass from perfectly indistinguishable (bosonic behaviour), to perfectly distinguishable (fermionic behavior), therefore demonstrating a mechanism for how spacetime can affect quantum systems. The work is increasingly relevant in the real world as we move towards global satellite quantum communications, and paves the way for further fundamental research that could test the influence of non-inertial motion (and equivalently curved spacetime) on quantum entanglement.

\end{abstract}

\maketitle
{\bf{Introduction.}}
Quantum field theory in curved spacetime, a theoretical framework for quantum behaviour in background gravitational fields, indicates that motion and underlying spacetime will have non-trivial effects on quantum systems. It has had success in predicting new quantum effects such as Hawking radiation \cite{hawkingBlack1974,hawkingParticle1975} and the Unruh effect \cite{fullingNonuniqueness1973,unruhNotes1976}. 
However, so far only a few of these new effects have been shown, and only in analogue systems \cite{Ulf,weinfurtnerMeasurement2011,steinhauerObservation2014,steinhauerObservation2016,munozdenovaObservation2019,Cromb2020,Braidotti2021}. 
Understanding all the effects that spacetime can have on quantum states is also becoming increasingly technically relevant as quantum communications aim towards satellite networks, which will have to account for the curvature of spacetime around the Earth \cite{kohlrusQuantum2017,barzelObserver2022}.

An improvement over analogue systems, but a so far rarely exploited experimental approach, relies on the equivalence of curved spacetime to accelerating and non-inertial frames. It is generally easier to create accelerations in laboratory embodiments than to do  space-based experiments, and allows access to regimes outside those found in our Solar System. Quantum technology is now sufficiently robust that we can start to test entanglement in various non-inertial frames. Fink et al. were able to place a bound on the (non-)effect of uniform acceleration on entanglement from 0.03g to 30g  with a drop tower as well as on a centrifuge \cite{finkExperimental2017}.

Non-inertial motion was also shown to influence the temporal distinguishability of photons by shifting the delay between them \cite{restucciaPhoton2019}, combining the relativistic Sagnac effect \cite{sagnacEther1913,postSagnac1967,ardittySagnac1981} with Hong-Ou-Mandel (HOM) interference  \cite{hongMeasurement1987}, and comparing the Sagnac effect between classical and quantum light. 

Here we report an experiment using non-inertial motion to alter the form of quantum entanglement between two photons. By altering the rotation speed of a modified HOM interferometer we are able to change a Hong-Ou-Mandel interference dip into a peak, antisymmetrising the entangled state and changing bosonic photon behaviour into  `fermionic' behaviour. Non-inertial motion therefore affects photon indistinguishability, putting to experimental test the proposed mechanism by which the Sagnac effect alters the symmetry of quantum entangled states \cite{torosRevealing2020}.

{\bf{Hong-Ou-Mandel Interference.}}
Hong-Ou-Mandel (HOM) biphoton interference \cite{hongMeasurement1987} provides information about the distinguishability of photons.
When two independent single photons cross at a lossless 50:50 beamsplitter, the unitarity of the beam splitter transformation, combined with the photon bosonic commutation relations, results in an interference forcing indistinguishable photons to `bunch’ and exit the beamsplitter through the same port. 
A time delay between the input photons creates distinguishability between the photons. Counting coincident detections between single photon detectors in the two output paths, a dip in the coincidence rate is observed when the photons temporally overlap. The visibility of the dip indicates overall indistinguishability in all photon properties.

In an analogous experiment with fermions, the fermionic anti-commutation relations would suppress the bunching of independent fermions and a peak in the output coincidences would be observed instead. This `fermionic' behaviour can also be observed with bosons  if the particles are entangled in an antisymmetric state \cite{wangTwophoton2003,wangQuantum2006,fedrizziAntisymmetrization2009}, which can be engineered in a number of ways \cite{pittmanCan1996,kwiatObservation1992,walbornMultimode2003,strekalovWhat1998,sagioroTime2004,olindoHongOuMandel2006,olindoErasing2015,abouraddyQuantumoptical2002,nasrDemonstration2003}.

{\bf{Outline of the experiment.}} 
\begin{figure}[t]
\includegraphics[width=\columnwidth]{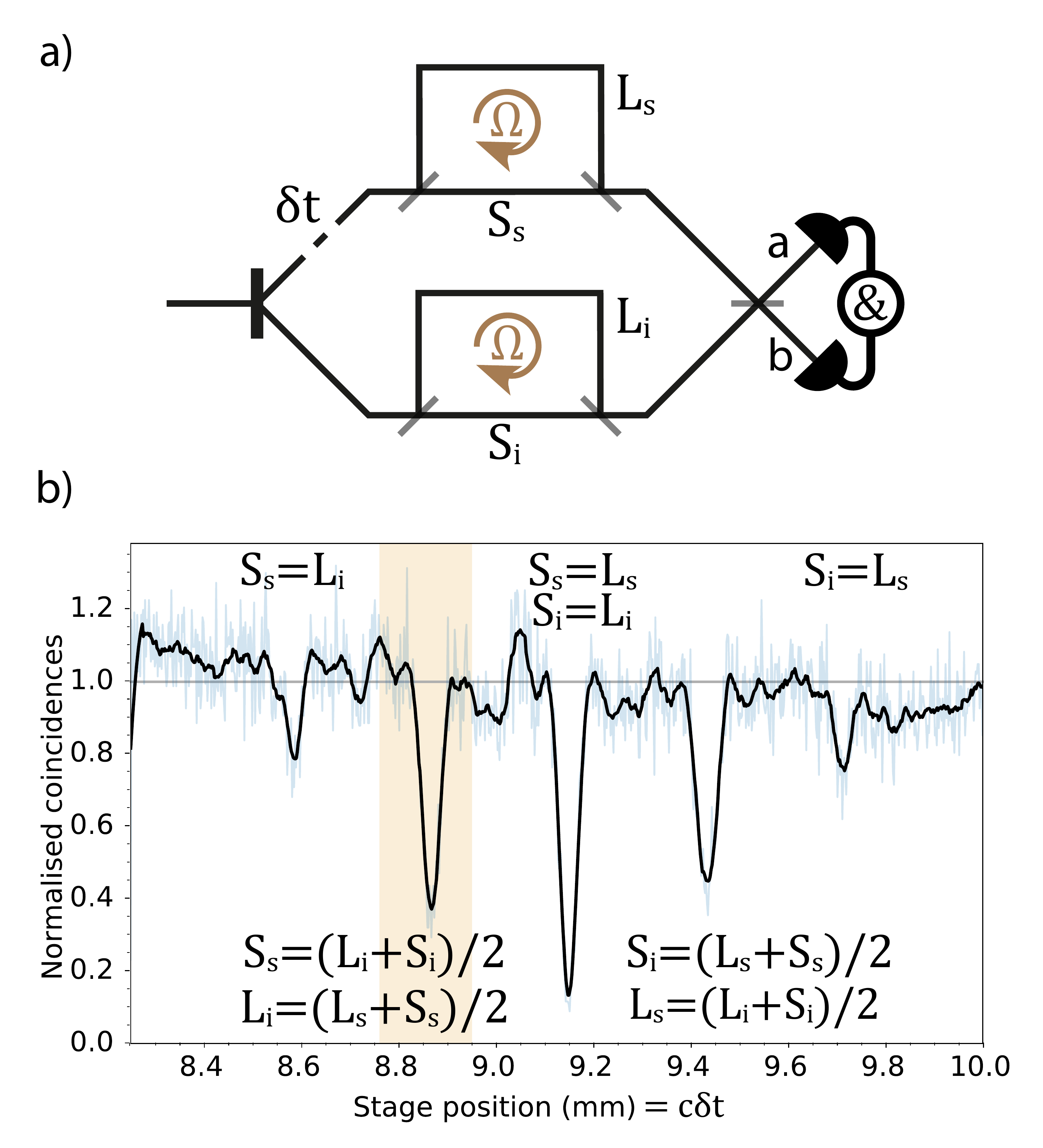}
\caption{{\bf{a) Schematic layout.}} Figurative diagram showing short $S$ and long $L$ path lengths (which travel against and with the rotation $\Omega$ direction) for the signal and idler photons in the system, along with the HOM delay $\delta t$ that scans the delay of one arm with respect to the other, and the detection in coincidence after the HOM beamsplitter. {\bf{b) Experimental scan of dips while not rotating.}} A graph of detected coincidences against the position of the stage, which is proportional to the HOM delay $\delta t$. Raw experimental data is shown in light blue, as well as a smoothed average in black. Five dips in the coincidences are present, corresponding to the different combinations of path lengths $S_s,L_i$ etc. at which HOM interference can occur. The shaded region shows an example range over which the stage is scanned when the experiment is in rotation.}
\label{f:schematic}
\end{figure}
%
%
A schematic of the rotating Hong-Ou-Mandel (HOM) interferometer is shown in Fig.~\ref{f:schematic}a. A pair of indistinguishable time-frequency entangled photons are produced in a nonlinear crystal and travel in separate arms (denoted with index `s', indicating the `signal' photon, and `i', indicating the `idler' photon) until interfering at a final beamsplitter, after which they are detected in coincidence. In each of the arms, each photon is also split 50:50 into two directions, taking either a long path ($L_{\{s,i\}}$) travelling clockwise, or a short path ($S_{\{s,i\}}$) travelling anticlockwise.
The extra paths are set so that the arms are symmetric, $L_s - S_s = L_i - S_i$. A variable overall delay $\delta t$ is also added into the signal arm, which varies both $S_s$ and $L_s$ equally. There are three different settings of the overall delay at which photons cross the beamsplitter at the same time: when $S_s=L_i$, when $L_s = S_i$, and when both $L_s = L_i, S_s = S_i$. The various combinations of the extra paths therefore result in additional HOM dips in the coincidence measurements at different delays $\delta t$. 

If the input light is entangled (rather than being two independent single photons), two additional interference features appear between these dips. These correspond to the delays at which $S_s= (L_i + S_i)/2, L_i = (L_s + S_s)/2$ and $L_s = (L_i + S_i)/2, S_i = (L_s+S_s)/2$. These additional interference features can be dips, but depending on the modulo $2\pi$ phase between paths $S$ and $L$, they can disappear completely, or can flip to become peaks~\cite{strekalovWhat1998,sagioroTime2004,olindoHongOuMandel2006,olindoErasing2015,abouraddyQuantumoptical2002,nasrDemonstration2003}.

The experiment is mounted on a rotating table. When the experiment is put into rotation at angular frequency $\Omega$, the Sagnac effect changes the time it takes for light to travel with, or against the rotation direction by
\begin{equation}\label{e:sagnac2}
    \Delta t_{\textrm{Sagnac}} =  \frac{4 A \Omega}{c^2}.
\end{equation}

Although the path lengths $S$ and $L$ are fixed, when rotating the Sagnac time delay changes the phase difference between them, scaling with the area $A$ enclosed by the paths. Increasing the rotation frequency so that the Sagnac phase difference between $S$ and $L$ paths increases by $\pi$ is therefore expected to flip these interferences from dips to peaks or vice versa, altering the entanglement symmetry and changing the indistinguishability of the photons as measured by the HOM - purely through non-inertial motion. 

\begin{figure}[t]
\includegraphics[width=0.6\columnwidth]{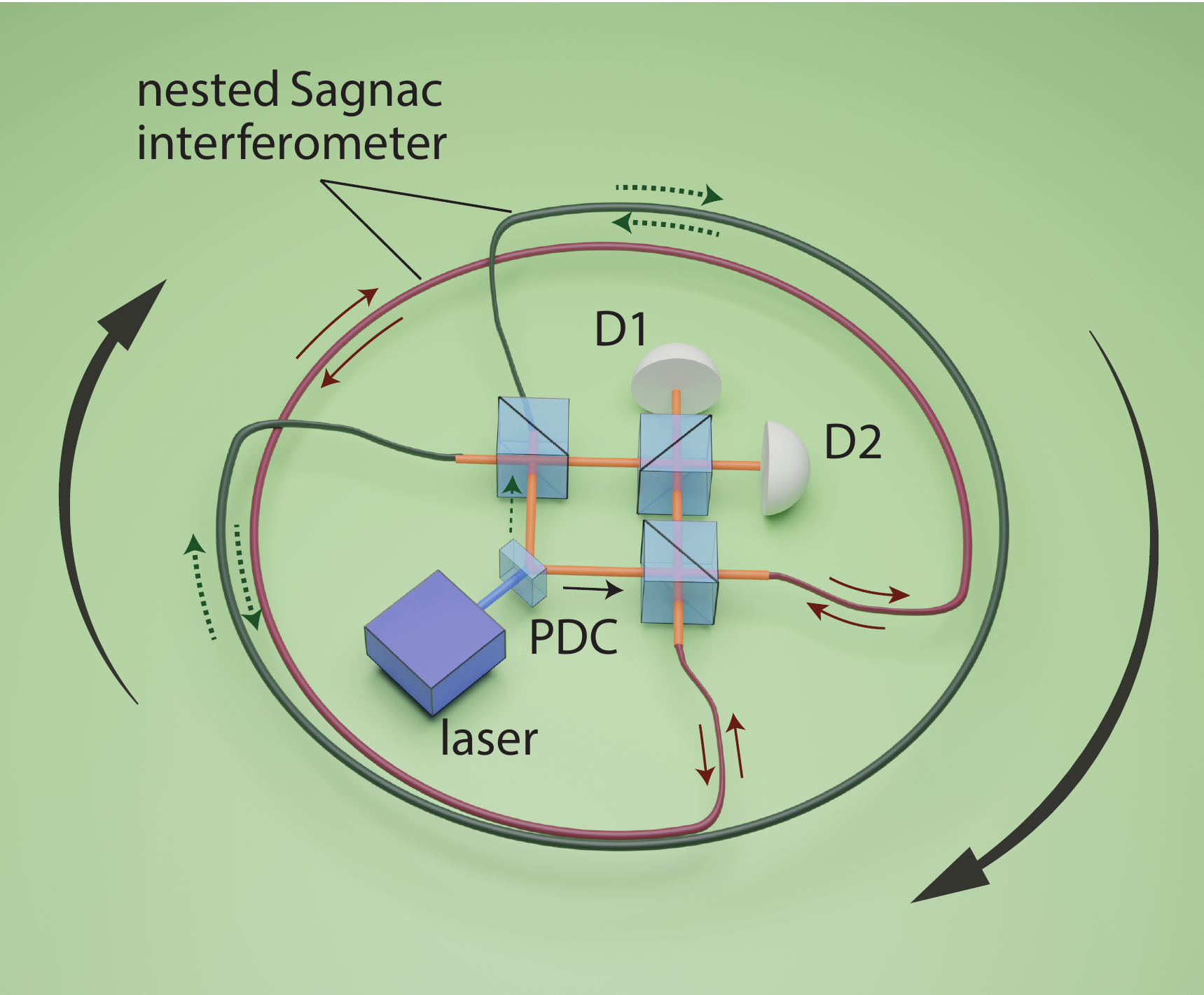}
\includegraphics[width=0.36\columnwidth]{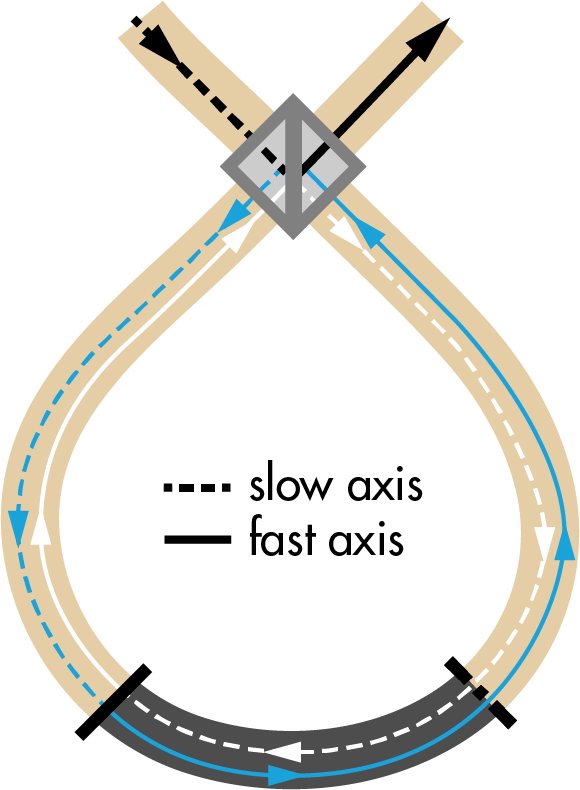}

\caption{{\bf{Diagram of experiment.}} Left) Rotating Hong-Ou-Mandel experiment with nested Sagnac interferometers that enclose the area of the rotating platform. A pump laser produces SPDC photon pairs at a nonlinear crystal which each pass through a Sagnac interferometer and then interfere at a beamsplitter. D1 and D2 are single photon detectors which can measure in coincidence.  Right) Detail of the 1 m birefringent delay between clockwise and anticlockwise directions created by the fibre loop in the nested Sagnac interferometers. This makes the HOM peaks easier to observe as a birefringent delay larger than the single photon coherence length ensures a consistent amount of light passing through the Sagnacs. Two 20 m polarisation-maintaining fibres are connected by a hybrid 1 m patch cord that flips the polarisation axis, so one direction in the Sagnac loop has 21 m of slow axis and 20 m of fast axis, the other has 21 m of fast axis and 20 m of slow axis.} 
\label{f:setup}
\end{figure}

{\bf{Theoretical model.}}
We follow a similar approach to Ref.~\cite{olindoHongOuMandel2006}. For our input state we assume degenerate Type I spontaneous parametric down-conversion (SPDC) pumped by frequency $\omega_p$ and add a variable delay $\delta t$ between the signal and idler photon arms: 
\begin{equation}\label{e:SPDC1}
    \ket{\psi} = \int_{0}^{\omega_p} d\omega B(\omega) e^{-i\omega\delta t} a_i^{\dagger}(\omega) a_s^{\dagger}(\omega_p - \omega) \ket{0},
\end{equation}
where $a_i^{\dagger}(\omega), a_s^{\dagger}(\omega_p - \omega)$ are the creation operators for modes of frequency $\omega, \omega_p - \omega$ for idler and signal photons, and $B(\omega)$ is the spectrum of the biphoton wavepacket.

Each arm contains a nested fibre Sagnac interferometer where the light hits a beamsplitter, splits into clockwise ($cw$) and anticlockwise ($ac$) directions, propagates in opposite directions through the same fibre loop for time $t_{\{cw,ac\}}$, and recombines when it hits the beamsplitter again upon exiting the nested Sagnac interferometer.
\begin{multline}
    a_i^{\dagger}(\omega) \mapsto \frac{1}{{2}} \left( e^{-i\omega t_{i,cw}} - e^{-i\omega t_{i,ac}} \right) a_{i,out}^{\dagger}(\omega) \\ + \frac{i}{{2}} \left( e^{-i\omega t_{i,cw}} + e^{-i\omega t_{i,ac}} \right) a_{i,back}^{\dagger}(\omega)
\end{multline}
As well as the Sagnac delay created between clockwise and anticlockwise photons travelling in a total fibre length $L_f$,
the polarisation maintaining fibre paths are constructed such that there is an additional constant birefringent delay from a mismatch between refractive indices $n_{cw}$ and $n_{ac}$ over a length $L_{b} \ll L_f$ (see Fig.~\ref{f:setup}b). This extra net delay is independent of rotation and creates the short $S$ and long $L$ paths in the Fig.~\ref{f:schematic}a schematic, ensuring separation into a total of five interference features shown in Fig~\ref{f:schematic}b. The total time delays are thus:
\begin{equation}
    \begin{split}
    t_{cw}(\Omega) &= \frac{L_{b} n_{cw}}{c} + \frac{L_f r \Omega}{c^2}\:,\\
    t_{ac}(\Omega) &= \frac{L_{b} n_{ac}}{c} - \frac{L_f r \Omega}{c^2}\:.
    \end{split}
\end{equation}
Here we assume that $L_f$, $L_b$ and $r$ are the same for signal and idler and thus $t_{i,cw}= t_{s,cw} = t_{cw}$ and $t_{i,ac}= t_{s,ac} = t_{ac}$ (for a more general approach, see Supplementary Material).

The light that exits the Sagnacs ($a_{i,out}, a_{s,out}$) interferes at the HOM beamsplitter, at which outputs ($a,b$) we find the final state:  
\begin{multline}
    \ket{\psi_{\textrm{final}}} = \frac{1}{8}\int_{0}^{\omega_p} d\omega B(\omega) e^{-i\omega\delta t}\\
    \left( e^{-i\omega t_{cw}} - e^{-i\omega t_{ac}} \right) \left( e^{-i(\omega_p-\omega)t_{cw}} - e^{-i (\omega_p-\omega) t_{ac}} \right)\\ \left( i a^{\dagger}(\omega) + b^{\dagger}(\omega)\right)  \left( a^{\dagger}(\omega_p-\omega) + i b^{\dagger}(\omega_p-\omega)\right) \ket{0}.
\end{multline}
The expected coincidences $N_c$, measured between two single photon detectors in the output arms, is calculated (details in Supplementary Material). 
Assuming a Gaussian spectrum for $B(\omega)$) of characteristic width $\Delta \omega$ we find
\begin{multline}\label{e:theoryfinalNc}
N_c  \propto 
C_b 
 -e^{-{\Delta\omega}^2 (\delta t + \Delta t)^2}
-e^{-{\Delta\omega}^2 (\delta t - \Delta t)^2}\\
+ 4 \cos{(\tfrac{\omega_p}{2} \Delta t)} \left( e^{-{{\Delta\omega}^2}(\delta t + \frac{\Delta t}{2})^2} + e^{-{{\Delta\omega}^2}(\delta t - \frac{\Delta t}{2})^2} \right) \\
-4e^{-{\Delta\omega}^2 \delta t^2} - 2\cos(\omega_p \Delta t)e^{-{{\Delta\omega}^2}\delta t^2}, 
\end{multline}
where $\Delta t = t_{cw}-t_{ac}$. Eq.~\eqref{e:theoryfinalNc} contains a term that does not depend on the HOM delay $\delta t$ and that forms the coincidence background: 
\begin{multline}\label{e:bg}
    C_b = 4 - 8e^{- \frac{{\Delta\omega}^2}{4} (\Delta t)^2 } \cos{\left(\frac{\omega_p}{2} \Delta t\right)}\\
    + 2 \cos{\left({\omega_p} \Delta t\right)}
    + 2e^{-{{\Delta\omega}^2} (\Delta t)^2 }.
\end{multline}
Of the terms in Eq.~\eqref{e:theoryfinalNc} that depend on $\delta t$ and describe interference features, three describe `fixed' HOM dips; a central dip and two smaller dips either side. There are then three oscillating terms, two which describe two fully oscillating dips/peaks in between the central `fixed' dip and side dips, and another which can increase the depth of the central dip (essentially ensuring the central dip remains fully visible when the light is fully indistinguishable even as the background $C_b$ fluctuates). From the periodicity of the fully oscillating dips (the $ \cos{(\tfrac{\omega_p}{2} \Delta t)}$ term) we find that a change in rotation frequency of ${c\lambda_p}/{(4 \pi L_f r)} $ Hz is required to fully flip a dip into a peak.

{\bf{Experimental Apparatus.}}
The experiment shown in Fig.~\ref{f:setup} is mounted on a rotating table driven by a stepper motor (RS-PRO, 180-5292) run by a controller module (Geckodrive, G201X). A UV pump laser (355 nm, Coherent Genesis CX STM) produces degenerate down-converted photon pairs ($\lambda$ = 710 nm) at a Type I BBO crystal. These (symmetrically) frequency-entangled photons are separated using a knife-edge prism, filtered (10 nm bandwidth), and each coupled into a polarisation maintaining fibre (PMF). One fibre coupler is mounted on a translation stage in order to scan the temporal delay $\delta t$.  Each fibre arm contains a nested Sagnac interferometer, consisting of a beamsplitter with its reflection and transmission ports connected by a 41 m loop of PMF. This optical fibre link is secured around the rotating platform in loops of diameter 0.908 m. The 41 m fibre link is made up of three fibre optic cables connected in series: two 20 m lengths with a 1 m fibre in the middle which has one key aligned to the slow axis and the other key aligned to the fast axis (shown in Fig.~\ref{f:setup}). This 1 m fibre flips the polarisation axis as the light travels around, creating a fixed net 1 m birefringent delay (beat length $\sim 1.1$ mm) between light travelling in different directions around the loop, creating short and long path options. 
As the two Sagnac interferometers do not share the same optical fibre, any temperature fluctuations affecting one fibre and not the other can introduce unwanted noise. To minimise these issues, the two Sagnac fibres are looped alongside each other and thermally insulated. 
After the Sagnac interferometers, the light in each arm recombines at a final HOM beamsplitter and the photons at the outputs are detected by single-photon avalanche diodes (SPADs) which measure singles and coincidences within a coincidence window of 5 ns. 

Measuring coincidences while scanning the delay $\delta t$ resulted in the series of five HOM dips (shown in Fig.~\ref{f:schematic}b) as expected from the different paths in the system and cross interference between the paths.
\begin{figure*}[t]
\includegraphics[width=1\linewidth]{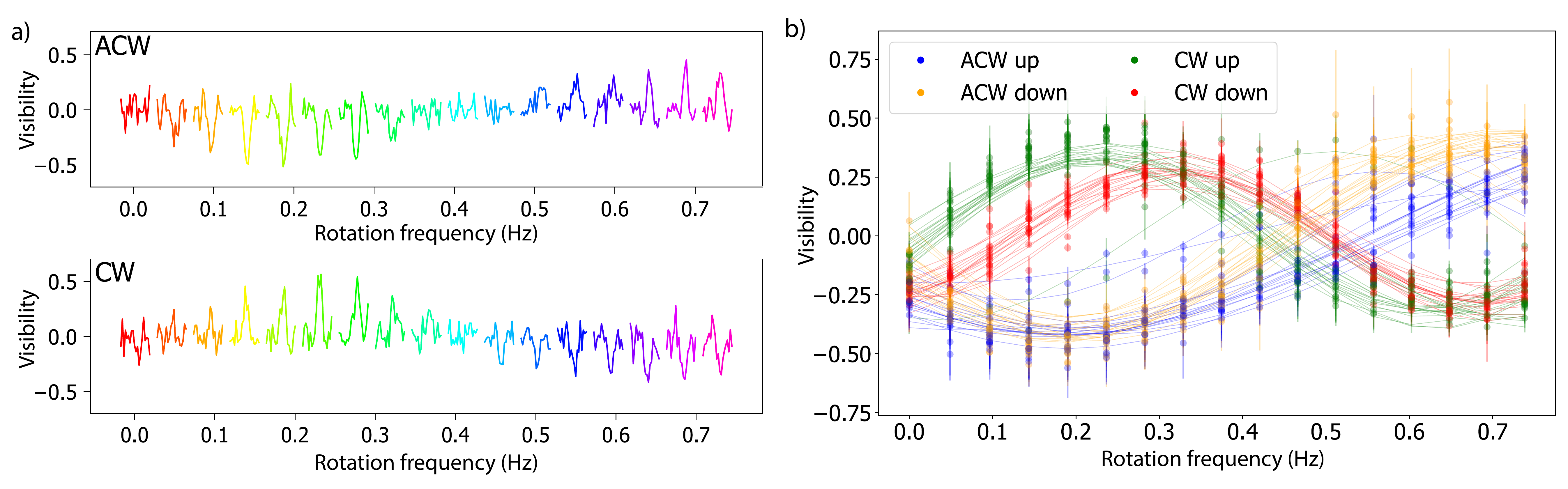}
\vspace*{-1em}\caption{{\bf{a) Dips inverting for clockwise and anticlockwise rotations.}} Each scan of the oscillating dip is taken at a set rotation frequency, the sequential discrete rotation frequencies used (from 0 to 0.735 Hz) are marked by colour. Contrasting adjacent clockwise and anticlockwise runs, to show how the effect depends on direction, consistent with the Sagnac effect. {\bf{b) Interference peak values with increasing rotation speed.}} Data from a long overnight measurement run ($\sim80$ sequences). Plots peak height over background for each rotation speed (dots) and the best fit sinusoids for each 0-0.74 Hz (`up') or 0.74-0 Hz (`down') sequence. Each sequence took $\sim10$ minutes. 
It is clear that clockwise and anticlockwise rotation move the peak in opposite directions from a similar starting phase. }
\label{f:dipflipCWACWall}
\end{figure*}

When the set-up is rotated at a constant speed, the Sagnac effect causes an additional phase shift between light travelling clockwise and anticlockwise around the loops, and with a large enough change in rotation speed this additional phase shift changes the symmetry of the entangled biphoton state such that the cross-interference features can flip from a dip to a peak and vice versa. 

The rotation speed of the apparatus was changed from 0 Hz to a maximum rotation speed ($\sim0.735$ Hz) in equally set steps (see in Supplementary Material for more information).
It was then stepped back down again to 0 Hz. These sequences were repeated, alternating between rotating anticlockwise and clockwise. The maximum speed was set conservatively to ensure the experiment could be repeated consistently over several hours without damage to the equipment or changes to the alignment due to vibrations at higher rotation speeds. 

At each rotation speed, the delay stage was used to scan over the second dip from the left (the shaded region in Fig.~\ref{f:schematic}b) in equal steps; the singles and coincidences were measured at each delay stage position for a short acquisition time. Most of the data was taken with a 10 $\mu$m step size and a 1.5 s acquisition time. Background coincidence rates for these measurements were of the order 100 counts/s.

{\bf{Results.}}
The results in Fig.~\ref{f:dipflipCWACWall} clearly show that the rotation changed the biphoton state as predicted and that the HOM interference changed smoothly and sinusoidally from a dip to a peak and vice versa as the rotation was stepped up or down.  

Depending on whether the experiment is spinning clockwise or anticlockwise, the Sagnac effect will either increase the phase between the nested paths or decrease it. As such, we expect that if we start from neither a dip nor a peak then rotating the experiment in one direction will turn it to a peak first as rotation speed increases, and the other direction will turn into a dip first as rotation speed increases. This can be seen in the experimental data (Fig.~\ref{f:dipflipCWACWall}). This dependence on rotation direction confirms that the main observed effect is due to the predicted Sagnac effect and is not due to spurious effects caused by centrifugal forces on the setup, which would not be dependent on the sense of rotation.

As the experiment consists of many metres of optical fibre, it was also sensitive to temperature changes \cite{shupeThermally1980} from the lab environment and from the operation of the electronics and the motor in the experiment. 
These temperature changes added extra phase drifts that changed over time, and thus also altered across measurement sequences precisely how many rotation steps were required to see a flip of the dip. To reduce these temperature noise effects, the measurements were performed in short time intervals whilst retaining an acceptable signal-to-noise ratio. 
We then averaged over 151 individual rotation sequences in order to average out small random changes and fluctuations in the environment. Some difference in periodicity might be anticipated between the clockwise and anticlockwise directions due to the g-force on the fibres mentioned above creating a common phase offset that in one direction works with, and in the other against, the Sagnac effect. Indeed, the mean of the 78 clockwise measurements was 0.41 Hz, and the mean of the 73 anticlockwise measurements was 0.53 Hz. 
Overall, averaging across all data, we measured a dip-to-peak rotation change (half period) of mean $0.47^{+0.10}_{-0.11}$ Hz, and median $0.43$ Hz, that matched well our theoretical expected value of 0.455 Hz.

{\bf{Conclusions.}}
We have shown that the statistics of biphoton interference can change depending on the non-inertial motion of the experimental frame. Non-inertial motion modifies the entanglement symmetry of the input biphoton state such that we observe Hong-Ou-Mandel interference dips (`bosonic' behaviour) change into peaks (`fermionic' behaviour) and vice versa, with changes in rotation speed of the set-up. This experimental change is consistent with the magnitude and directionality of the Sagnac effect mechanism at the heart of our theoretical model.   

The dips that show this change do not appear in our simulations if we use two independent identical single photons as input; the mechanism for changing photon statistics acts on the frequency correlations between the photons that arise from the time-frequency entanglement of the photon pair. 

We live in a rotating frame here on Earth, with its influence on the quantum mechanical phase of single particles already recorded \cite{wernerEffect1979,gustavsonPrecision1997}. The fact that the underlying spacetime can have effects on photon entanglement may also have consequences for quantum communication technologies, particularly over long distances and using satellites. 

This work shows the promising utility of combining photonic technologies and non-inertial motion for testing fundamental physics questions at the interface of quantum mechanics and general relativity. Taking these ideas and techniques further, it could be possible to create entanglement with rotational motion \cite{torosGeneration2022} or with other forms of non-inertial motion, particularly with non-uniform acceleration as indicated by theoretical research into quantum field theories in curved spacetimes \cite{bruschiVoyage2012, friisMotion2012, alsingObserver2012}.

{\bf{Acknowledgements.}} 
The authors acknowledge financial support from the Leverhulme Trust, the Royal Academy of Engineering Chairs in Emerging Technology, the Royal Society Professorship scheme and the UK Engineering and Physical Sciences Research Council (grant no. EP/W007444/1).\\ 

\bibliographystyle{apsrev4-2} 
\bibliography{rotflipHOM3.bib}



\newpage
\onecolumngrid

\section{On the distinguishability of photons in a rotating reference frame: Supplementary Information}

\section{Extended Theoretical Model}\label{as:theory}

In this section we present a detailed theoretical analysis of the interferometer.
We first define the initial state (Sec.~\eqref{subsec:Initial-biphoton-state}),
followed by the general analysis of the interference when we have asymmetry between the arms (Sec.~\eqref{subsec:derivation}),
and finally show that the coincidence probability, in the considered
experimental regime, is not affected by the biphoton width (Sec.~\ref{subsec:Biphoton-frequency-spread}).

\subsection{Initial biphoton state\label{subsec:Initial-biphoton-state}}

For our input state we assume degenerate Type I SPDC pumped by frequency
$\omega_{p}$. We represent a single signal photon in one arm and
a single idler photon in the other arm as creation mode operators
$a_{s}^{\dagger},a_{i}^{\dagger}$ acting on the vacuum
$\ket{0}$. The initial state is given by

\begin{equation}
\vert\psi_{\text{i}}\rangle=\int d\omega_{1}d\omega_{2}\psi_{\text{i}}(\omega_{1},\omega_{2})a_{i}^{\dagger}(\omega_{1})a_{s}^{\dagger}(\omega_{2})\vert0\rangle,\label{eq:initial}
\end{equation}
the biphoton spectrum is 
\begin{alignat}{1}
\psi_{\text{i}}(\omega_{1},\omega_{2})=\frac{1}{\mathcal{N}_{\text{i}}} & e^{-\frac{(\mu-\text{\ensuremath{\omega}}_{1})^{2}}{4\Delta\omega^{2}}}e^{-\frac{(\mu-\text{\ensuremath{\omega}}_{2})^{2}}{4\Delta\omega^{2}}}e^{-\frac{(\text{\ensuremath{\omega}}_{1}+\text{\ensuremath{\omega}}_{2}-2\mu)^{2}}{4\text{\ensuremath{\sigma_{p}}}^{2}}},
\end{alignat}
and the normalization is
\begin{equation}
\mathcal{N}_{\text{i}}=\sqrt{\frac{2\pi\Delta\omega}{\sqrt{\frac{1}{\Delta\omega^{2}}+\frac{2}{\sigma_{p}^{2}}}}}.\label{eq:norm}
\end{equation}
The behaviour of two-photon interference depends on three parameters
of the initial state: the mean photon frequency $\mu=\omega_{p}/2$,
the single photon frequency spread $\Delta\omega$, and the biphoton
frequency spread $\sigma_{p}$. 

We now first discuss in detail the case $\sigma_{p}\ll\Delta\omega$, 
providing a full derivation (section \ref{subsec:derivation}).
We will then validate this approximation by considering the more general analysis with a
finite $\sigma_{p}$ which can be performed using similar steps (section
\ref{subsec:Biphoton-frequency-spread}).

\subsection{Derivation of the number of coincidences\label{subsec:derivation}}

In this section we present a more general and
detailed treatment of the interferometer, allowing for asymmetry between
the arms. In the experiment, fluctuation in the middle dip - even
flipping as well - was sometimes observed, which can only be explained
by this more general model, caused by an imbalance between the birefringent
delays in one arm compared to the other, exacerbated and changed by
noise. 

To obtain a simple expression with a Gaussian spectrum we divide 
Eq.~\eqref{eq:initial} by $ \sqrt{4\pi \sigma_p} \sqrt[4]{2 \sigma ^2+\sigma_p^2}$ to find:
\begin{equation}
\vert\psi\rangle  =\int d\omega_{1}d\omega_{2}\left[\frac{1}{\sqrt{2\pi(\sqrt{2}\sigma_{p})^{2}}}e^{-\frac{(\text{\ensuremath{\omega}}_{1}+\text{\ensuremath{\omega}}_{2}-2\mu)^{2}}{2(\sqrt{2}\text{\ensuremath{\sigma_{p}}})^{2}}}\right]
  \left[\frac{1}{\sqrt{2\pi \sigma^2}} e^{-\frac{(\mu-\text{\ensuremath{\omega}}_{1})^{2}}{4\Delta\omega^{2}}}e^{-\frac{(\mu-\text{\ensuremath{\omega}}_{2})^{2}}{4\Delta\omega^{2}}}\right]a_{i}^{\dagger}(\omega_{1})a_{s}^{\dagger}(\omega_{2})\vert0\rangle.\label{eq:approximating}
\end{equation}
The term in the first square bracket of Eq.~\eqref{eq:approximating}
is a Gaussian which can be approximated with a Dirac delta function
$\delta(\text{\ensuremath{\omega}}_{1}+\text{\ensuremath{\omega}}_{2}-2\mu)$ in the limit $\sigma_{p}\rightarrow0$ (see Sec.~\ref{subsec:Biphoton-frequency-spread} for the general model with a finite $\sigma_p$).
Performing the integration over $\omega_{2}$,
and relabelling $\omega_{1}\rightarrow\omega$, we obtain: 
\begin{equation}
\ket{\psi}=\int_{0}^{\omega_{p}}d\omega B(\omega)a_{i}^{\dagger}(\omega)a_{s}^{\dagger}(\omega_{p}-\omega)\ket{0},\label{e:SPDC}
\end{equation}
where the spectrum of the biphoton wavepacket $B(\omega)$ is a gaussian
centred at $\mu=\omega_{p}/2$ and frequency spread $\Delta\omega$ (which depends on the filters used in the experiment). We now follow
analogous steps as discussed in Ref~\citep{olindoHongOuMandel2006},
applied to the experimental setting presented in the main text.

We include a variable delay $\delta t$ between the signal and idler photon arms, to represent the scannable HOM delay:
\begin{equation}
    \ket{\psi} = \int_{0}^{\omega_p} d\omega B(\omega) e^{-i\omega\delta t} a_i^{\dagger}(\omega) a_s^{\dagger}(\omega_p - \omega) \ket{0}.
\end{equation}

Each arm contains a nested fibre Sagnac interferometer, formed of a beamsplitter with the reflection and transmission ports connected by looped optical fibre. Upon entering the Sagnac interferometer, the mode splits into clockwise (cw) and anticlockwise (ac) directions:
\begin{equation}
  a_i^{\dagger}(\omega) \mapsto \frac{1}{\sqrt{2}} \left( a_{i,cw}^{\dagger}(\omega) + i a_{i,ac}^{\dagger}(\omega)\right).
\end{equation}
Afterwards, these modes propagate in opposite directions through the same fibre for a time $t_{cw,ac}$ and so pick up a phase $\phi_{cw}(\omega) = \omega t_{cw}(\Omega,n)$ that changes with rotation speed $\Omega$:
\begin{equation}\label{e:SagBSphase}
a_i^{\dagger}(\omega) \mapsto \frac{1}{\sqrt{2}} \left(e^{-i\phi_{i,cw}} a_{i,cw}^{\dagger}(\omega) + e^{-i\phi_{i,ac}} i a_{i,ac}^{\dagger}(\omega)\right).
\end{equation}

As well as the Sagnac delay created between clockwise and anticlockwise photons travelling in a total fibre length $L_f$ looped in radius $r$ on a platform rotating clockwise at $\Omega$, the fibre paths are constructed such that there is an additional constant birefringent delay from a mismatch between refractive indices $n_{cw}$ and $n_{ac}$ over a length $L_{b} << L_f$:  
\begin{equation}
    \begin{split}
    t_{cw} &= \frac{L_{b} n_{cw}}{c} + \frac{L_f r \Omega}{c^2},\\
    t_{ac} &= \frac{L_{b} n_{ac}}{c} - \frac{L_f r \Omega}{c^2}.
    \end{split}
\end{equation}
Here we continue with a more  experimentally-realistic generality that any or all of $L_f$, $L_b$ and $r$ could be slightly different for signal and idler, and thus in general $t_{i,cw} \neq t_{s,cw}$ etc. 

The clockwise and anticlockwise paths interfere as they pass the beamsplitter again:
\begin{equation}\label{e:SagBSout}
    \begin{split}
    a_{i,cw}^{\dagger}(\omega) &\mapsto \frac{1}{\sqrt{2}} \left( a_{i,out}^{\dagger}(\omega) + i a_{i,back}^{\dagger}(\omega)\right),\\
    a_{i,ac}^{\dagger}(\omega) &\mapsto \frac{1}{\sqrt{2}} \left( i a_{i,out}^{\dagger}(\omega) + a_{i,back}^{\dagger}(\omega)\right).
    \end{split}
\end{equation}
Combining Eqs. \eqref{e:SagBSphase} and \eqref{e:SagBSout} gives:
\begin{equation}
    a_i^{\dagger}(\omega) \mapsto \frac{1}{{2}} \left( e^{-i\phi_{i,cw}(\omega)} - e^{-i\phi_{i,ac}(\omega)} \right) a_{i,out}^{\dagger}(\omega) + \frac{i}{{2}} \left( e^{-i\phi_{i,cw}(\omega)} + e^{-i\phi_{i,ac}(\omega)} \right) a_{i,back}^{\dagger}(\omega),
\end{equation}
and similarly for the signal mode we find:
\begin{multline}
    a_s^{\dagger}(\omega_p-\omega) \mapsto \frac{1}{{2}} \left( e^{-i\phi_{s,cw}(\omega_p-\omega)} - e^{-i\phi_{s,ac}(\omega_p-\omega)} \right) a_{s,out}^{\dagger}(\omega_p-\omega)\\ + \frac{i}{{2}} \left( e^{-i\phi_{s,cw}(\omega_p-\omega)} + e^{-i\phi_{s,ac}(\omega_p-\omega)} \right) a_{s,back}^{\dagger}(\omega_p-\omega).
\end{multline}
We only consider light that exits the Sagnac towards the HOM beamsplitter ($a_{i,out}, a_{s,out}$): 
\begin{multline}\label{e:sagnacsout}
    \ket{\psi_{\textrm{Sagnacs}}} = \frac{1}{4}\int_{0}^{\omega_p} d\omega B(\omega) e^{-i\omega\delta t}  \left( e^{-i\phi_{i,cw}(\omega)} - e^{-i\phi_{i,ac}(\omega)} \right) a_{i,out}^{\dagger}(\omega) \\
    \left( e^{-i\phi_{s,cw}(\omega_p-\omega)} - e^{-i\phi_{s,ac}(\omega_p-\omega)} \right) a_{s,out}^{\dagger}(\omega_p-\omega) \ket{0}. 
\end{multline}
At the HOM beamsplitter there is the last mode transformation (input ($a_{i,out}, a_{s,out}$), output ($a,b$)): 
\begin{align}\label{e:HOMbstransform}
        a_{i,out}^{\dagger} &\mapsto \frac{1}{\sqrt{2}} \left( i a^{\dagger} + b^{\dagger}\right) &
        a_{s,out}^{\dagger} &\mapsto \frac{1}{\sqrt{2}} \left( a^{\dagger} + i b^{\dagger}\right).
\end{align}
Then from Eqs. \eqref{e:sagnacsout} and \eqref{e:HOMbstransform} we finally find: 
\begin{multline}
    \ket{\psi_{\textrm{final}}} = \frac{1}{8}\int_{0}^{\omega_p} d\omega B(\omega) e^{-i\omega\delta t}  \left( e^{-i\phi_{i,cw}(\omega)} - e^{-i\phi_{i,ac}(\omega)} \right) \left( e^{-i\phi_{s,cw}(\omega_p-\omega)} - e^{-i\phi_{s,ac}(\omega_p-\omega)} \right)\\ \left( i a^{\dagger}(\omega) + b^{\dagger}(\omega)\right)  \left( a^{\dagger}(\omega_p-\omega) + i b^{\dagger}(\omega_p-\omega)\right) \ket{0}.
\end{multline}

For the state of the electromagnetic field exiting the beamsplitter, we want to find the expected number of coincidences $N_c$. A coincidence detection is detecting one photon at one detector, and another photon at the other detector within a small, finite coincidence window $\tau_f$. Specifically, we define the coincidence probability as the probability of detecting a photon in detector $a$ at time $t$ and a photon in detector $b$ at time $t + \tau$ is given by $P(\tau)$. 

\begin{equation}\label{e:NCPtau}
    N_c = \int_{-\tau_f}^{\tau_f} d\tau P(\tau),
\end{equation}
\begin{equation}\label{e:probabilitytau}
    P(\tau)  = \bra{\psi_{\textrm{final}}} E_a^- (t) E_b^- (t+\tau) E_b^+ (t+\tau) E_a^+ (t) \ket{\psi_{\textrm{final}}},
\end{equation}
    \begin{align}
        E_a^+ (t) &= \int d\omega e^{-i\omega t} a(\omega),  &        E_b^+ (t) &= \int d\omega e^{-i\omega t} b(\omega). 
    \end{align}

Let us now evaluate Eq.~\eqref{e:probabilitytau}. We first calculate $E_b^+ (t+\tau) E_a^+ (t) \ket{\psi_{\textrm{final}}}$:
\begin{multline}\label{e:EbEaP}
    E_b^+ (t+\tau) E_a^+ (t) \ket{\psi_{\textrm{final}}} = 
    \frac{1}{8} \int d\omega_2 \int d\omega_1 \int_0^{\omega_p} d\omega e^{-i\omega_2 (t + \tau)} b(\omega_2)
     e^{-i\omega_1 t} a(\omega_1) e^{-i\omega \delta t} B(\omega) \\ \left( e^{-i\phi_{i,cw}(\omega)} - e^{-i\phi_{i,ac}(\omega)} \right) \left( e^{-i\phi_{s,cw}(\omega_p-\omega)} - e^{-i\phi_{s,ac}(\omega_p-\omega)} \right)\\
    \left( i  a^{\dagger}(\omega)  a^{\dagger}(\omega_p-\omega) - a^{\dagger}(\omega) b^{\dagger}(\omega_p-\omega) + b^{\dagger}(\omega)  a^{\dagger}(\omega_p-\omega) + i b^{\dagger}(\omega)  b^{\dagger}(\omega_p-\omega) \right) \ket{0}.
\end{multline}
The $a^{\dagger}(\omega)  a^{\dagger}(\omega_p-\omega)$ and $b^{\dagger}(\omega)  b^{\dagger}(\omega_p-\omega)$ terms are bunching terms (not coincidences) and give a contribution of zero. The only non-zero contributions to the coincidence count comes from the $a^{\dagger}(\omega) b^{\dagger}(\omega_p-\omega)$ and $b^{\dagger}(\omega)  a^{\dagger}(\omega_p-\omega)$ terms when $\omega_1 = \omega,\omega_2 = \omega_p - \omega$ or when $\omega_1 = \omega_p - \omega,\omega_2 = \omega$. Eq.~\eqref{e:EbEaP} thus simplifies to: 
\begin{multline}\label{e:EbEafinal1}
    E_b^+ (t+\tau) E_a^+ (t) \ket{\psi_{\textrm{final}}} = \\
    \frac{1}{8} \int_0^{\omega_p} d\omega e^{-i\omega \delta t} B(\omega)  \left( e^{-i\phi_{i,cw}(\omega)} - e^{-i\phi_{i,ac}(\omega)} \right) \left( e^{-i\phi_{s,cw}(\omega_p-\omega)} - e^{-i\phi_{s,ac}(\omega_p-\omega)} \right)\\
    \left( e^{-i\omega(t+\tau)} e^{-i(\omega_p-\omega)t} - e^{-i(\omega_p-\omega)(t+\tau)} e^{-i\omega t}\right) \left(1 + \delta(\omega_p-2\omega)\right) \ket{0}. 
\end{multline}
The $\delta(\omega_p-2\omega)$ term in Eq.~\eqref{e:EbEafinal1} evaluates to zero: 
\begin{multline}
\frac{1}{8} \int_0^{\omega_p} d\omega e^{-i\omega \delta t} B(\omega)  \left( e^{-i\phi_{i,cw}(\omega)} - e^{-i\phi_{i,ac}(\omega)} \right) \left( e^{-i\phi_{s,cw}(\omega_p-\omega)} - e^{-i\phi_{s,ac}(\omega_p-\omega)} \right)\\ \hfill \left(e^{-i(\omega_p-\omega)(t+\tau)} e^{-i\omega t} - e^{-i\omega(t+\tau)} e^{-i(\omega_p-\omega)t} \right) \delta(\omega_p-2\omega)  \ket{0} \\
= \frac{1}{8} e^{-i\frac{\omega_p}{2} \delta t} B(\frac{\omega_p}{2} )  \left( e^{-i\phi_{i,cw}(\frac{\omega_p}{2} )} - e^{-i\phi_{i,ac}(\frac{\omega_p}{2} )} \right) \left( e^{-i\phi_{s,cw}(\frac{\omega_p}{2} )} - e^{-i\phi_{s,ac}(\frac{\omega_p}{2} )} \right)\\ \left(e^{-i(\frac{\omega_p}{2} )(t+\tau)} e^{-i\frac{\omega_p}{2}  t} - e^{-i\frac{\omega_p}{2} (t+\tau)} e^{-i(\frac{\omega_p}{2} )t} \right) \ket{0}
=0.
\end{multline}
The global phase ($e^{-i\omega_p t}$) in the remaining term of Eq.~\eqref{e:EbEafinal1} can be factored out, resulting in: 
\begin{multline}\label{e:moveglobph}
E_b^+ (t+\tau) E_a^+ (t) \ket{\psi_{final}} = 
\frac{1}{8} e^{-i\omega_p t} \int_0^{\omega_p} d\omega e^{-i\omega \delta t} B(\omega)  \left( e^{-i\phi_{i,cw}(\omega)} - e^{-i\phi_{i,ac}(\omega)} \right) \\ \hfill \left( e^{-i\phi_{s,cw}(\omega_p-\omega)} - e^{-i\phi_{s,ac}(\omega_p-\omega)} \right)
\left( e^{-i\omega(t+\tau)} e^{+i\omega t} - e^{-i\omega_p \tau}e^{+i\omega(t+\tau)} e^{-i\omega t}\right) \ket{0} \\
= 
\frac{1}{8} e^{-i\omega_p t} \int_0^{\omega_p} d\omega e^{-i\omega \delta t} B(\omega)  \left( e^{-i\phi_{i,cw}(\omega)} - e^{-i\phi_{i,ac}(\omega)} \right) \\ \left( e^{-i\phi_{s,cw}(\omega_p-\omega)} - e^{-i\phi_{s,ac}(\omega_p-\omega)} \right) \left( e^{-i\omega\tau} - e^{-i\omega_p\tau}e^{+i\omega\tau}\right) \ket{0}.
\end{multline}

We do a change of variables  $\omega_{} \mapsto (\omega + \omega_p/2)$; $\omega = \omega_{} - \omega_p/2$, and factor out global phases that won't contribute to the final probability. From Eq.~\eqref{e:moveglobph} we thus find:
%
\begin{multline}\label{e:changedvars}
    E_b^+ (t+\tau) E_a^+ (t) \ket{\psi_{final}} = 
    \frac{1}{8} e^{-i\omega_p t} e^{-i\frac{\omega_p}{2} \delta t} e^{-i\frac{\omega_p}{2}\tau} \int_{-\omega_p/2}^{\omega_p/2} d\omega e^{-i\omega \delta t} B(\omega + \omega_p/2)  \left( e^{-i\phi_{i,cw}(\omega + \omega_p/2)} - e^{-i\phi_{i,ac}(\omega + \omega_p/2)} \right)\\
    \left( e^{-i\phi_{s,cw}(\omega_p/2-\omega)} - e^{-i\phi_{s,ac}(\omega_p/2-\omega )} \right) \left( e^{-i\omega\tau} - e^{+i\omega\tau}\right) \ket{0}.
\end{multline}

Multiplying Eq.~\eqref{e:changedvars} by the conjugate transpose to get $P(\tau)$ we thus find:
\begin{multline}\label{e:ptau}
    P(\tau) = \frac{1}{16} \int_{-\omega_p/2}^{\omega_p/2} d\omega \int_{-\omega_p/2}^{\omega_p/2} d\omega' e^{+i\omega' \delta t} e^{-i\omega \delta t}  B^*(\omega' + \omega_p/2) B(\omega + \omega_p/2)
    \left( e^{+i\omega'\tau} - e^{-i\omega'\tau}\right) \left( e^{-i\omega\tau} - e^{+i\omega\tau}\right)\\ \left( e^{+i\phi_{i,cw}(\omega' + \omega_p/2)} - e^{+i\phi_{i,ac}(\omega' + \omega_p/2)} \right) \left( e^{-i\phi_{i,cw}(\omega + \omega_p/2)} - e^{-i\phi_{i,ac}(\omega + \omega_p/2)} \right)\\
    \left( e^{+i\phi_{s,cw}(\omega_p/2-\omega')} - e^{+i\phi_{s,ac}(\omega_p/2-\omega' )} \right) \left( e^{-i\phi_{s,cw}(\omega_p/2-\omega)} - e^{-i\phi_{s,ac}(\omega_p/2-\omega )} \right).
\end{multline}

We now insert Eq.~\eqref{e:ptau} into Eq.~\eqref{e:NCPtau} to obtain the number of coincidences. The integration over $\tau$ can be simplified by assuming $\tau$ is the longest timescale in the system, so that the limits can be taken to infinity and we can use $\int_{-\infty}^{\infty} d\tau e^{i(\omega_1+\omega_2)\tau} = 2\pi\delta{(\omega_1+\omega_2)}$. Hence the time integration can be carried out analytically and we find:
\begin{equation}
\begin{aligned}
    \int_{-\infty}^{\infty} d\tau \left(e^{+i\omega'\tau} - e^{-i\omega'\tau}\right) \left( e^{-i\omega\tau} - e^{+i\omega\tau}\right)
    &= 4\pi (\delta(\omega' -\omega) - \delta(\omega'+\omega)).
\end{aligned}
\end{equation}
The expression for the number of coincidences thus reduces to:
\begin{multline}
    N_c = \frac{4\pi}{16} \int_{-\omega_p/2}^{\omega_p/2} d\omega 
    \biggr[ e^{+i\omega \delta t}  e^{-i\omega \delta t}  B^*(\omega + \omega_p/2) B(\omega + \omega_p/2)\\
    \left( e^{+i\phi_{i,cw}(\omega + \omega_p/2)} - e^{+i\phi_{i,ac}(\omega + \omega_p/2)} \right) \left( e^{-i\phi_{i,cw}(\omega + \omega_p/2)} - e^{-i\phi_{i,ac}(\omega + \omega_p/2)} \right)\\
    \left( e^{+i\phi_{s,cw}(\omega_p/2-\omega)} - e^{+i\phi_{s,ac}(\omega_p/2-\omega )} \right) \left( e^{-i\phi_{s,cw}(\omega_p/2-\omega)} - e^{-i\phi_{s,ac}(\omega_p/2-\omega )} \right) \\
    - e^{-i2\omega \delta t}  B^*(-\omega + \omega_p/2) B(\omega + \omega_p/2) \left( e^{+i\phi_{i,cw}(-\omega + \omega_p/2)} - e^{+i\phi_{i,ac}(-\omega + \omega_p/2)} \right) \left( e^{-i\phi_{i,cw}(\omega + \omega_p/2)} - e^{-i\phi_{i,ac}(\omega + \omega_p/2)} \right)\\
    \left( e^{+i\phi_{s,cw}(\omega_p/2+\omega)} - e^{+i\phi_{s,ac}(\omega_p/2+\omega )} \right) \left( e^{-i\phi_{s,cw}(\omega_p/2-\omega)} - e^{-i\phi_{s,ac}(\omega_p/2-\omega )} \right) \biggr].
\end{multline}

Assuming symmetry of $B$ around $\omega_p/2$ we see the emergence of two parts, one that does not depend on the HOM delay $\delta t$ (the coincidence background), and one that does, that gives rise to dips (or peaks) at specific delays:
\begin{multline}\label{e:NCphis}
    N_c = \frac{4\pi}{16} \int_{-\omega_p/2}^{\omega_p/2} d\omega  {|B(\omega + \omega_p/2)|^2}  \biggr[  \left( e^{+i\phi_{i,cw}(\omega + \omega_p/2)} - e^{+i\phi_{i,ac}(\omega + \omega_p/2)} \right) \left( e^{-i\phi_{i,cw}(\omega + \omega_p/2)} - e^{-i\phi_{i,ac}(\omega + \omega_p/2)} \right)\\
    \left( e^{+i\phi_{s,cw}(\omega_p/2-\omega)} - e^{+i\phi_{s,ac}(\omega_p/2-\omega )} \right) \left( e^{-i\phi_{s,cw}(\omega_p/2-\omega)} - e^{-i\phi_{s,ac}(\omega_p/2-\omega )} \right) \\
    - e^{-i 2 \omega \delta t} \left( e^{+i\phi_{i,cw}(-\omega + \omega_p/2)} - e^{+i\phi_{i,ac}(-\omega + \omega_p/2)} \right) \left( e^{-i\phi_{i,cw}(\omega + \omega_p/2)} - e^{-i\phi_{i,ac}(\omega + \omega_p/2)} \right)\\
    \left( e^{+i\phi_{s,cw}(\omega_p/2+\omega)} - e^{+i\phi_{s,ac}(\omega_p/2+\omega )} \right) \left( e^{-i\phi_{s,cw}(\omega_p/2-\omega)} - e^{-i\phi_{s,ac}(\omega_p/2-\omega )} \right) \biggr].
\end{multline}

From Eq.~\eqref{e:NCphis} using the linearity of the phase shifts $\phi_{cw}(\omega) = \omega t_{cw}(\Omega,n), \phi_{ac}(\omega) = \omega t_{ac}(\Omega,n)$, we find:
%
\begin{multline}\label{e:NCtime}
N_c = \frac{4\pi}{16} \int_{-\omega_p/2}^{\omega_p/2} d\omega  {|B(\omega + \omega_p/2)|^2} \biggr[
\left( 2 - 2\cos\left((\omega + \omega_p/2)(t_{i,cw}-t_{i,ac})\right) \right) \left( 2-2\cos\left((\omega_p/2 - \omega) (t_{s,cw} - t_{s,ac})\right) \right) \\
- e^{-i 2 \omega \delta t} \left( e^{-i2\omega t_{i,cw}} + e^{-i 2\omega t_{i,ac}} - 2 \cos{\left(\tfrac{\omega_p}{2} ( t_{i,cw}-t_{i,ac})\right)} e^{-i\omega(t_{i,cw} + t_{i,ac})}  \right) \\
\left( e^{+i 2\omega t_{s,cw}} + e^{+i 2\omega t_{s,ac}} - 2 \cos{\left(\tfrac{\omega_p}{2} (t_{s,cw} - t_{s,ac})\right)} e^{+i\omega(t_{s,cw} + t_{s,ac})}\right) \biggr].
\end{multline}

We now insert the Gaussian for $B(\omega + \omega_p/2)$, and in addition assuming the Gaussian spread $\delta\omega << \frac{\omega_p}{2}$ we can extend the integration limits to infinity. Eq.~\eqref{e:NCtime} simplifies to:
\begin{multline}\label{e:NCgaussian}
N_c = \frac{4\pi}{16} \int_{-\infty}^{\infty} d\omega \frac{1}{2\pi(\Delta\omega)^2} e^{-\frac{\omega^2}{(\Delta\omega)^2}} \biggr[ 
4 \left( 1 - \cos\left((\omega + \omega_p/2)(t_{i,cw}-t_{i,ac})\right) \right) \left( 1 -\cos\left((\omega_p/2 - \omega) (t_{s,cw} - t_{s,ac})\right) \right) \\
- e^{-i 2 \omega \delta t} \left( e^{-i2\omega t_{i,cw}} + e^{-i 2\omega t_{i,ac}} - 2 \cos{\left(\tfrac{\omega_p}{2} ( t_{i,cw}-t_{i,ac})\right)} e^{-i\omega(t_{i,cw} + t_{i,ac})}  \right) \\
\left( e^{+i 2\omega t_{s,cw}} + e^{+i 2\omega t_{s,ac}} - 2 \cos{\left(\tfrac{\omega_p}{2} (t_{s,cw} - t_{s,ac})\right)} e^{+i\omega(t_{s,cw} + t_{s,ac})}\right) \biggr].
\end{multline}

We can evaluate the integral in Eq.~\eqref{e:NCgaussian} using
\begin{align}
    \int_{-\infty}^{\infty} d\omega e^{-a\omega^2 + b\omega + c} &= \sqrt{\frac{\pi}{a}} e^{\frac{b^2}{4a} + c}, \label{e:integrals1} \\
    \int_{-\infty}^{\infty} d\omega e^{-a\omega^2 + ik\omega} &= \sqrt{\frac{\pi}{a}} e^{-\frac{k^2}{4a}} \label{e:integrals2}, \\ \textrm{and}  \;
    \int_{-\infty}^{\infty} d\omega \cos{(a\omega)} &= Re\left[\int_{-\infty}^{\infty} d\omega e^{ia\omega} \right]. \label{e:integrals3}
\end{align}

Using Eqs.~\eqref{e:integrals1} - \eqref{e:integrals3} in Eq.~\eqref{e:NCgaussian} we find an analytic expression for the coincidences $N_c$:
\begin{multline}\label{e:NCfinalcomplex}
N_c  = \frac{\sqrt{\pi}}{8 (\Delta\omega)} \biggr[ 
\biggr( 4 - 4e^{- \frac{{\Delta\omega}^2}{4} (t_{iac}-t_{icw})^2 } \cos{\left(\frac{\omega_p}{2} (t_{iac}-t_{icw})\right)} 
- 4e^{-\frac{{\Delta\omega}^2}{4} (t_{sac}-t_{scw})^2 } \cos{\left(\frac{\omega_p}{2} (t_{sac}-t_{scw})\right)}\\
+ 2e^{-\frac{{\Delta\omega}^2}{4} (t_{iac}-t_{icw}-t_{sac}+t_{scw})^2 } \cos{\left(\frac{\omega_p}{2} (t_{iac}-t_{icw} + t_{sac}-t_{scw})\right)}\\
+ 2e^{-\frac{{\Delta\omega}^2}{4} (t_{iac}-t_{icw}+t_{sac}-t_{scw})^2 } \cos{\left(\frac{\omega_p}{2} (t_{iac}-t_{icw} - t_{sac}+t_{scw})\right)}
 \biggr)\\
\qquad + \biggr( 
-e^{-{\Delta\omega}^2 (\delta t + t_{icw}-t_{scw})^2} -e^{-{\Delta\omega}^2 (\delta t + t_{icw}-t_{sac})^2}
-e^{-{\Delta\omega}^2 (\delta t + t_{iac}-t_{scw})^2} -e^{-{\Delta\omega}^2 (\delta t + t_{iac}-t_{sac})^2}\\
+ 2 \cos{(\tfrac{\omega_p}{2} (t_{scw}-t_{sac}))} \left( e^{-\frac{{\Delta\omega}^2}{4}(2\delta t + 2t_{icw} -t_{scw} -t_{sac})^2} + e^{-\frac{{\Delta\omega}^2}{4}(2\delta t + 2t_{iac} -t_{scw} -t_{sac})^2} \right) \\
+ 2 \cos{(\tfrac{\omega_p}{2} (t_{icw}-t_{iac}))} \left( e^{-\frac{{\Delta\omega}^2}{4}(2\delta t - 2t_{scw} +t_{icw} + t_{iac})^2} + e^{-\frac{{\Delta\omega}^2}{4}(2\delta t - 2t_{sac} + t_{icw} +t_{iac})^2} \right) \\
- 4 \cos{(\tfrac{\omega_p}{2} (t_{scw}-t_{sac}))} \cos{(\tfrac{\omega_p}{2} (t_{icw}-t_{iac}))} e^{-\frac{{\Delta\omega}^2}{4}(2\delta t + t_{icw} +t_{iac} -t_{scw} -t_{sac} )^2} 
 \biggr) \biggr].
\end{multline}

In an ideal experiment, to simplify the effect, the differences between corresponding signal and idler paths should be zero, i.e. $t_{sac} = t_{iac} = t_{ac}$ and $t_{scw} = t_{icw} = t_{cw}$. Which further simplifies Eq.~\eqref{e:NCfinalcomplex} to:  
%
\begin{multline}
N_c  = \frac{\sqrt{\pi}}{8 (\Delta\omega)} \biggr[ 
\biggr( 4 - 8e^{- \frac{{\Delta\omega}^2}{4} (t_{ac}-t_{cw})^2 } \cos{\left(\frac{\omega_p}{2} (t_{ac}-t_{cw})\right)}
+ 2 \cos{\left({\omega_p} (t_{ac}-t_{cw})\right)}
+ 2e^{-{{\Delta\omega}^2} (t_{ac}-t_{cw})^2 }
 \biggr)\\
+ \biggr( 
-2e^{-{\Delta\omega}^2 \delta t^2} -e^{-{\Delta\omega}^2 (\delta t + t_{cw}-t_{ac})^2}
-e^{-{\Delta\omega}^2 (\delta t + t_{ac}-t_{cw})^2}\\
+ 4 \cos{(\tfrac{\omega_p}{2} (t_{ac}-t_{cw}))} \left( e^{-{{\Delta\omega}^2}(\delta t - \frac{t_{ac} -t_{cw}}{2})^2} + e^{-{{\Delta\omega}^2}(\delta t + \frac{t_{ac}-t_{cw}}{2})^2} \right) \\
- 4 \cos^2{(\tfrac{\omega_p}{2} (t_{ac}-t_{cw}))} e^{-{{\Delta\omega}^2}\delta t^2}  \biggr) \biggr]. \label{eq:Nctemp}
\end{multline}

\begin{figure}[h]
\includegraphics[width=0.68\columnwidth]{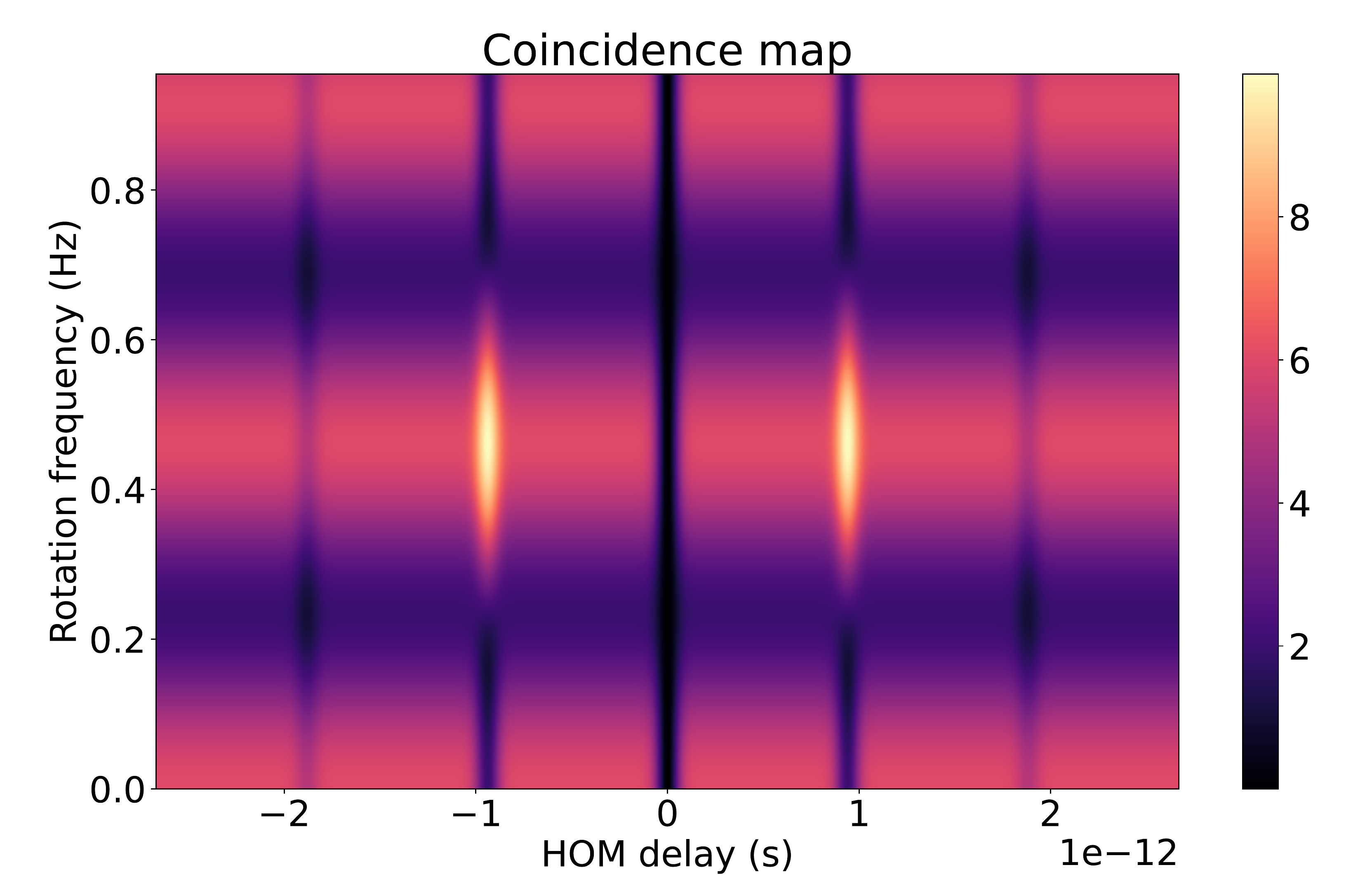}
\caption[Simulation results]{{\bf{Simulation results.}} Plots Equation \eqref{e:theoryfinal} using $\lambda_p = 355$nm, $L_f = 41$m, $r=0.454$m, $L_b=1$m, $\Delta \omega = 1.19\times 10^{13}$, $n_{cw}-n_{ac} = 5.641\times 10^{-4}$. 
Shows how coincidences are expected to change with rotation and delay. The changing dips go from maximum visibility dips (with respect to the background) to maximum visibility peaks with a 0.455 Hz rotation change. The background also changes with rotation.}%
\label{f:sim}
\end{figure}

To reach the form in the main paper, we can designate the background coincidence level $C_b$: 
\begin{equation}\label{e:bgSupp}
    C_b = 4 - 8e^{- \frac{{\Delta\omega}^2}{4} \Delta t^2 } \cos{\left(\frac{\omega_p}{2} \Delta t\right)}
    + 2 \cos{\left({\omega_p} \Delta t\right)}
    + 2e^{-{{\Delta\omega}^2} \Delta t^2 },
\end{equation}
where we have defined $\Delta t=t_{cw}-t_{ac}$. In addition we re-write the middle dip terms to isolate the oscillatory part:
\begin{equation}
    4 \cos^2{(\tfrac{\omega_p}{2} \Delta t)} e^{-{{\Delta\omega}^2}\delta t^2} = 2\cos(\omega_p \Delta t)e^{-{{\Delta\omega}^2}\delta t^2} + 2 e^{-{{\Delta\omega}^2}\delta t^2}.
\end{equation}
From Eq.~\eqref{eq:Nctemp} we then find: 
\begin{multline}\label{e:theoryfinal}
N_c  = \frac{\sqrt{\pi}}{8 \Delta\omega} \biggr[ 
C_b 
 -e^{-{\Delta\omega}^2 (\delta t + \Delta t)^2}
-e^{-{\Delta\omega}^2 (\delta t - \Delta t)^2}\\
+ 4 \cos{(\tfrac{\omega_p}{2} \Delta t)} \left( e^{-{{\Delta\omega}^2}(\delta t + \frac{\Delta t}{2})^2} + e^{-{{\Delta\omega}^2}(\delta t - \frac{\Delta t}{2})^2} \right)
-4e^{-{\Delta\omega}^2 \delta t^2} - 2\cos(\omega_p \Delta t)e^{-{{\Delta\omega}^2}\delta t^2} 
\biggr],
\end{multline}
which matches Eq.~(6) in the main text. The coincidence landscape plotted from the simulation, where the arms are symmetric (Eq.~\ref{e:theoryfinal}), is shown in Figure \ref{f:sim}.

\subsection{Effect of biphoton frequency spread\label{subsec:Biphoton-frequency-spread}}

In this section we discuss the general case with a finite
biphoton spread $\sigma_{p}$. The analysis is analogous to the one
discussed in detail in Sec.~\ref{subsec:derivation} but due to the significantly
longer expressions we report only the final results.

The final state (i.e., the state at the input of the final beamsplitter)
is given by

\begin{equation}
\vert\psi_{\text{f}}\rangle=\int d\omega_{1}d\omega_{2}\psi_{\text{f}}(\omega_{1},\omega_{2})\hat{a}^{\dagger}(\omega_{1})\hat{b}^{\dagger}(\omega_{2})\vert0\rangle, \label{eq:finalS}
\end{equation}
where the biphoton spectrum is

\begin{equation}
\psi_{\text{f}}(\omega_{1},\omega_{2})=  \frac{1}{\mathcal{N}_{\text{f}}}e^{-\frac{(\mu-\text{\ensuremath{\omega}}_{1})^{2}}{4\Delta\omega^{2}}}e^{-\frac{(\mu-\text{\ensuremath{\omega}}_{2})^{2}}{4\Delta\omega^{2}}}e^{-\frac{(\text{\ensuremath{\omega}}_{1}+\text{\ensuremath{\omega}}_{2}-2\mu)^{2}}{4\text{\ensuremath{\sigma_{p}}}^{2}}}
  \sin^{2}\left(\frac{\text{\ensuremath{\omega}}_{1}\Delta t}{2}\right)\sin^{2}\left(\frac{\text{\ensuremath{\omega_{2}}}\Delta t}{2}\right),\label{eq:spectrumf}
\end{equation}
and the normalization is given by

\begin{equation}
\mathcal{N_{\text{f}}}=\frac{\pi\Delta\omega^{2}\text{\ensuremath{\sigma_{p}}}e^{-\Delta\omega^{2}\Delta t^{2}}}{\sqrt{2\Delta\omega^{2}+\sigma_{p}^{2}}}\big( 1 -4\cos(\mu\Delta t)\exp\left(\frac{\Delta\omega^{2}\left(3\Delta\omega^{2}+\sigma_{p}^{2}\right)\Delta t^{2}}{2\left(2\Delta\omega^{2}+\sigma_{p}^{2}\right)}\right)
  +\cos(2\mu\Delta t)e^{\frac{2\Delta\omega^{4}\Delta t^{2}}{2\Delta\omega^{2}+\sigma_{p}^{2}}}+2e^{\Delta\omega^{2}\Delta t^{2}}\big).
\end{equation}
In the first part of Eq.~\eqref{eq:spectrumf} we recognize the initial
state defined in Eq.~\eqref{eq:initial}, while the last two factors contain
the interference contribution due to the imbalanced paths. 

The probability of preparing the final state in Eq.~\eqref{eq:finalS} is given by:
\begin{alignat}{1}
P_{\text{f}}=\frac{1}{8}\big[2 & -4\cos(\mu \Delta t)\exp\left(\frac{\Delta\omega^{2}\left(3\Delta\omega^{2}+\sigma_{p}^{2}\right)\Delta t^{2}}{2\left(2\Delta\omega^{2}+\sigma_{p}^{2}\right)}-\sigma^{2}\Delta t^{2}\right)\nonumber  \\ &+\cos(2\mu \Delta t)\exp\left(\frac{2\Delta\omega^{4}\Delta t^{2}}{2\Delta\omega^{2}+\sigma_{p}^{2}}-\Delta\omega^{2}\Delta t^{2}\right)+e^{-\Delta\omega^{2}\Delta t^{2}}\big],\label{eq:stateprobability}
\end{alignat}
as only the input modes of the final beamsplitter contribute to the final coincidence probability (for more details see the derivation leading up to Eq.~\eqref{e:sagnacsout}).

The coincidence probability can be computed using the formula~\cite{wangQuantum2006}:

\begin{equation}
P_{\text{c}}=\frac{1}{2}-\frac{1}{2}\int\int\psi_\text{f}^{*}(\omega_{1},\omega_{2})\psi_\text{f}(\omega_{2},\omega_{1})d\omega_{1}d\omega_{2}.\label{eq:generalPc}
\end{equation}
Inserting Eqs.~\eqref{eq:spectrumf} in \eqref{eq:generalPc} and
performing the integrations we find:

\begin{equation}
P_{\text{c}}(\delta t,\Delta t,\mu,\Delta\omega,\ensuremath{\sigma}_{p})=\frac{1}{2}\left(1-\frac{I_{1}+I_{2}+I_{3}+I_{4}}{S}\right),\label{eq:coincidenceprobability}
\end{equation}
where the interference is quantified by 

\begin{alignat}{1}
I_{1} & =4e^{-\Delta\omega^{2}(\ensuremath{\delta t}+\Delta t)(\text{\ensuremath{\delta t}}-\Delta t)}+e^{-\text{\ensuremath{\delta t}}\Delta\omega^{2}(\text{\ensuremath{\delta t}}+2\Delta t)}+e^{-\text{\ensuremath{\delta t}}\Delta\omega^{2}(\delta t-2\Delta t)},\\
I_{2} & =-4\cos(\mu\Delta t)\text{exp}\left(\frac{\Delta\omega^{2}\ensuremath{\sigma}_{p}^{2}\left(-2\text{\ensuremath{\delta t}}^{2}-2\delta t\Delta t+\Delta t^{2}\right)+\Delta\omega^{4}(-2\ensuremath{\delta t}-3\Delta t)(2\ensuremath{\delta t}-\Delta t)}{2\left(2\Delta\omega^{2}+\ensuremath{\sigma}_{p}^{2}\right)}\right),\\
I_{3} & =-4\cos(\mu\Delta t)\text{exp}\left(\frac{\Delta\omega^{2}\ensuremath{\sigma}_{p}^{2}\left(-2\text{\text{\ensuremath{\delta t}}}^{2}+\Delta t (\Delta t+2\delta t)\right)+\Delta\omega^{4}(-2\delta t-\Delta t)(2\delta t-3\Delta t)}{2\left(2\Delta\omega^{2}+\ensuremath{\sigma}_{p}^{2}\right)}\right),\\
I_{4} & =2\cos(2\mu\Delta t)\text{exp}\left(\frac{2\Delta\omega^{4}\Delta t^{2}}{2\Delta\omega^{2}+\ensuremath{\sigma}_{p}^{2}}-\delta t^{2}\Delta\omega^{2}\right),
\end{alignat}
and

\begin{equation}
S=2-8\cos(\mu\Delta t)\exp\left(\frac{\Delta\omega^{2}\left(3\Delta\omega^{2}+\ensuremath{\sigma}_{p}^{2}\right)\Delta t^{2}}{2\left(2\Delta\omega^{2}+\ensuremath{\sigma}_{p}^{2}\right)}\right)
  +2\cos(2\mu\Delta t)e^{\frac{2\Delta\omega^{4}\Delta t^{2}}{2\Delta\omega^{2}+\ensuremath{\sigma}_{p}^{2}}}+4e^{\Delta\omega^{2}\Delta t^{2}}.
\end{equation}

The number of counts (with unit incoming photon rate) is thus given by: 
\begin{equation}
N_{c}=P_{f}P_{c},\label{eq:Nc}
\end{equation}
where $P_{\text{f}}$ and $P_{\text{c}}$ are given in Eqs.~\eqref{eq:stateprobability}
and \eqref{eq:coincidenceprobability}, respectively. 
We recover Eq.~\eqref{e:theoryfinal} from Eq.~\eqref{eq:Nc} by multiplying it with the normalization factor $4\sqrt{\pi}/\Delta\omega$
and taking the limit $\sigma_{p}\rightarrow0$. 

We find that when $\sigma_{p}/2\pi \leq  2 \times 10^{10}$  Hz, which is the experimental parameter, the model predicts that the coincidence count $N_c$ is not significantly different from the one found for the case $\sigma_{p}\rightarrow 0$. On the other hand, by increasing the value of $\sigma_{p}$ to values close to the single-photon frequency spread $\Delta\omega$, the model predicts a loss of visibility of the two invertible HOM dips, which arise due the entangled nature of the biphoton state, as the degree of entanglement starts to decrease, as well as we start to move out of the biphoton coherence time.

\section{Rotation speed calibration}\label{as:rotcalib}
\begin{figure}[h!]
\includegraphics[width=0.6\columnwidth]{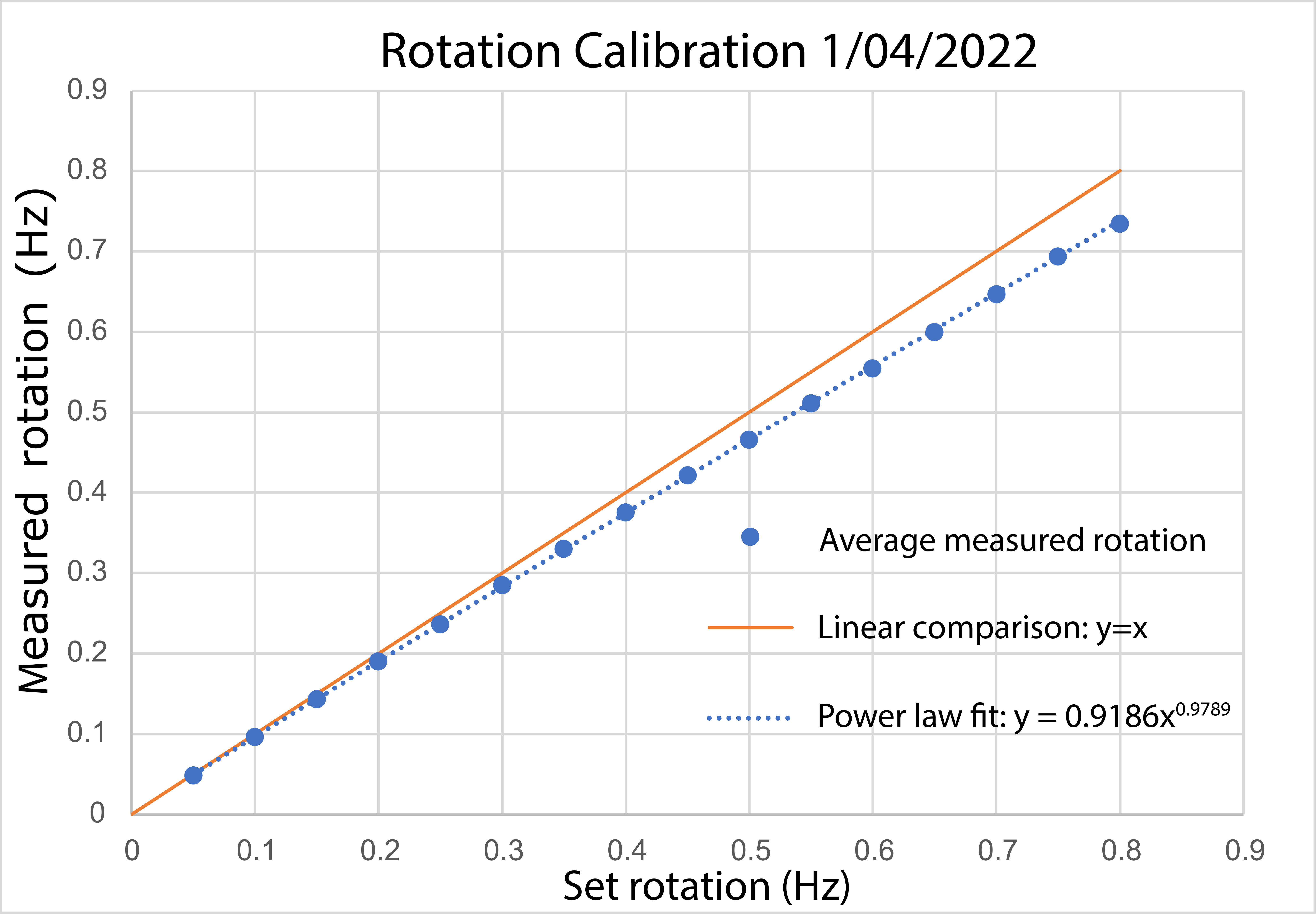}
\caption[Calibration of rotation frequency]{{\bf{Calibration of rotation frequency.}} The dotted line is a power law fit to the measured rotation. Theoretical linear relationship is shown for comparison. }
\label{f:rotcalib}
\end{figure}

The rotation frequency that the motor was set to was not necessarily the actual rotation speed of the experimental turntable due to friction effects. To quantify this and calibrate the measurements accordingly, the time it took to complete a certain number of revolutions at each rotation step used in the experiment was recorded for both clockwise and anticlockwise directions, and used to estimate the real rotation frequency of the experiment. There was no significant difference in magnitude between clockwise and anticlockwise directions for the same set rotation. Both clockwise and anticlockwise measurements were averaged and that data is shown in Fig.~\ref{f:rotcalib}, with a power law fit to the data and the ideal linear case for comparison. 

\pagebreak[3]
\section{Histogram of measurements}

\begin{figure}[h]
    \includegraphics[width=8cm]{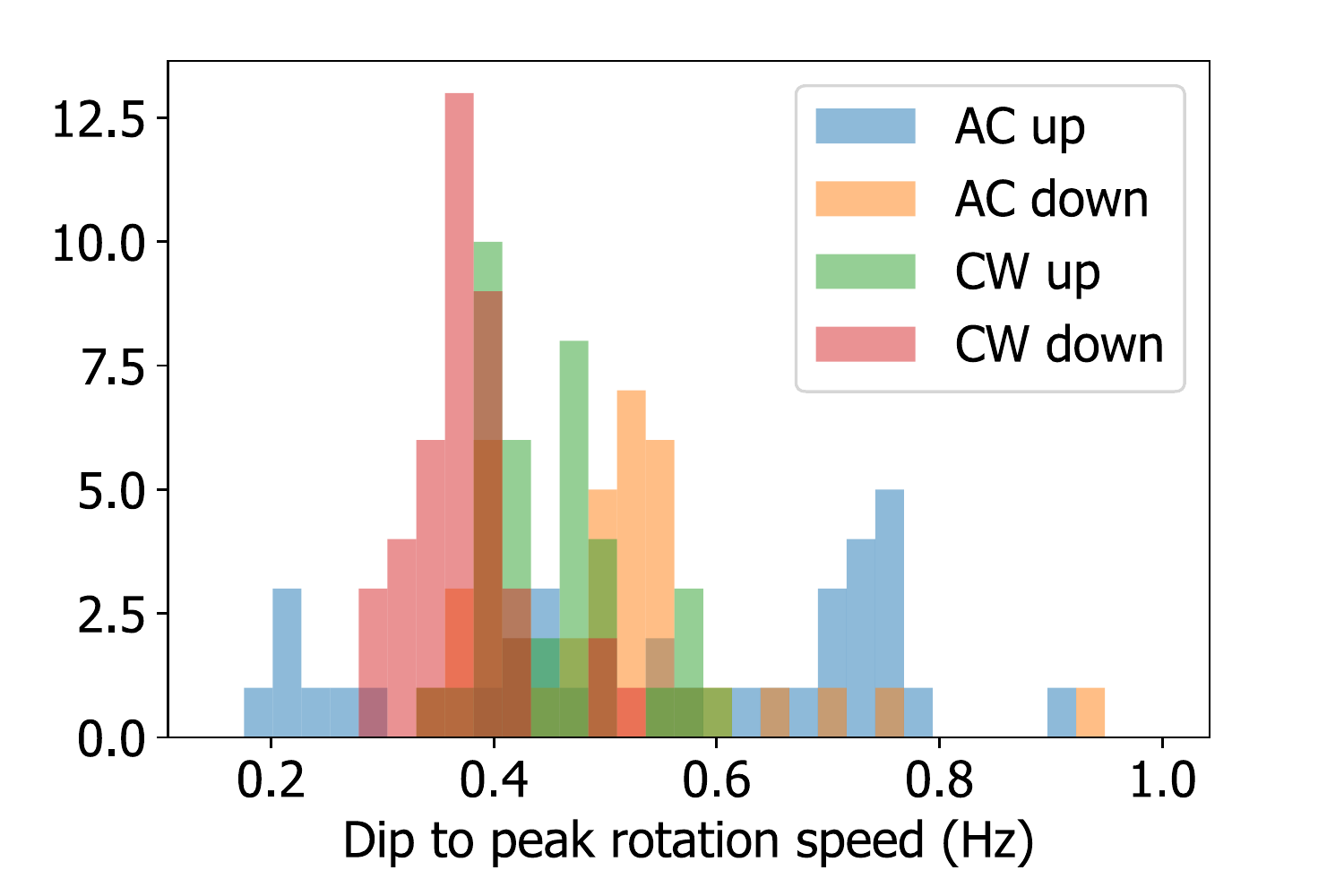} \includegraphics[width=8cm]{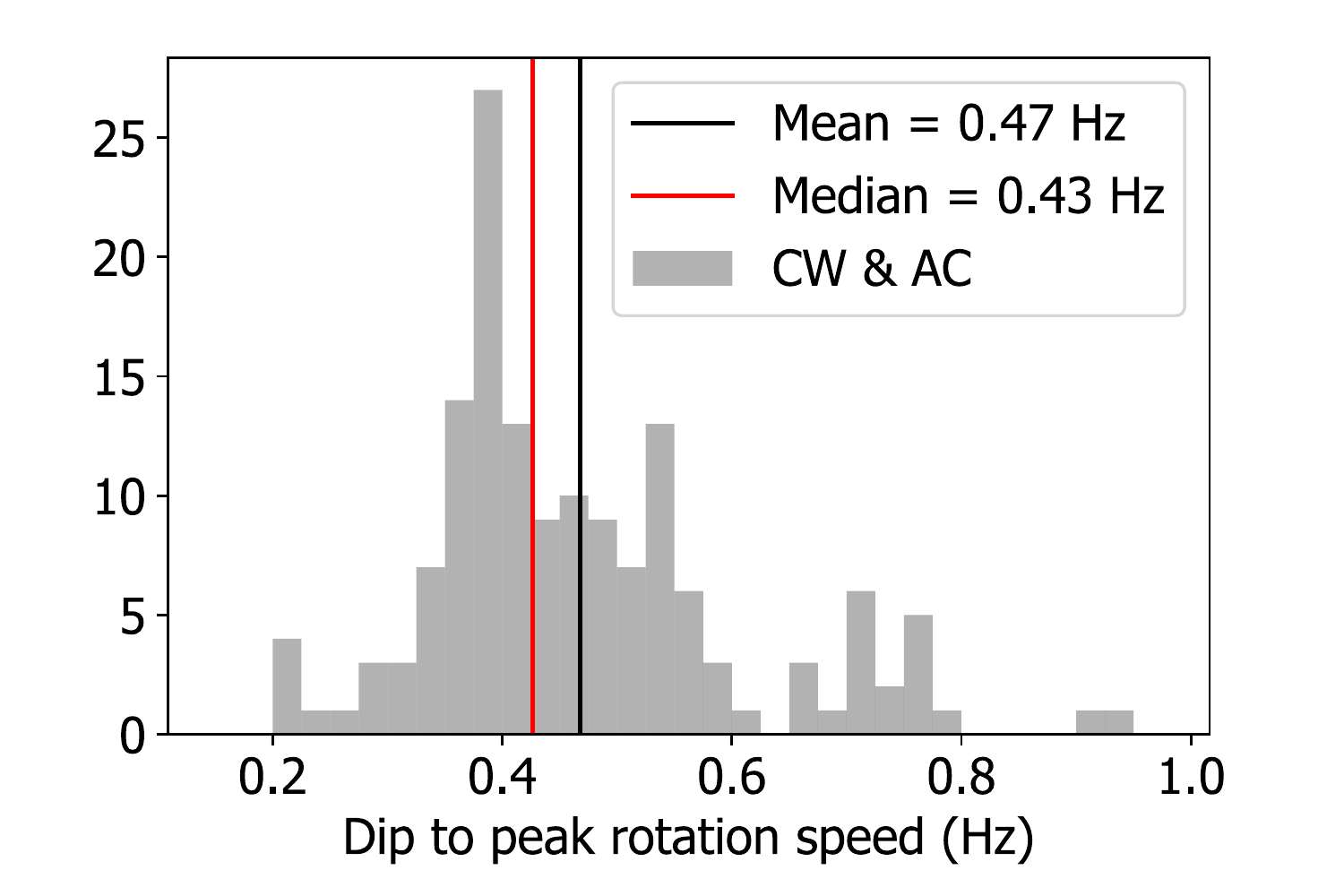}

\caption[Histograms showing rotation speed change required to fully flip from dip to peak (or vice versa)]{{\bf{Histograms showing best fit rotation speed change required to fully flip from dip to peak (or vice versa).}} Data extracted from sinusoidal fits to 151 (0-0.735) Hz sequences. Sorted into sequence type (left hand figure), and also shown overall (right hand figure) with the mean and median of the set indicated. 
}
\label{f:rot-phase hist}
\end{figure}

It was expected for the oscillating dip to have a period of $\frac{2 \pi c^2}{\omega_p L_f r}$ with respect to the angular frequency of $\Omega$ of the platform, with a corresponding change in rotation of $\frac{c\lambda_p}{4 \pi L_f r} $Hz required to fully flip a dip into a peak or vice versa. To find the change in rotation that caused a flip from a dip into a peak 
in our experiment, the amplitude of the centre of the peak (or dip) above the background was identified for each rotation speed setting within a sequence of 0 Hz-0.735 Hz (stepping either up or down). A sinusoid was then fit to the peak amplitude-rotation data, and if the fitting process converged well, the best fit parameter of the period of the curve was extracted. The half-periods (representing the rotation frequency change required to change a dip to a peak or vice versa) extracted from the data analysis are shown in Fig.~\ref{f:rot-phase hist}.

Best-fit periods were extracted from 78 clockwise rotation sequences, and 73 anti-clockwise rotation sequences. The mean and the median from the clockwise data, the anticlockwise data, and the whole set are shown in Table~\ref{t:meanmedian}. The mean and median of the whole set are also shown superimposed in Fig.~\ref{f:rot-phase hist}.

\begin{table}[h]
    \begin{tabular}{c|c|c|c}
        & Clockwise (78 runs) & Anticlockwise (73 runs) & Total (151 runs) \\
         \hline
        Mean & 0.411 & 0.528 & 0.468 \\
        Median & 0.396 & 0.530 & 0.427 
    \end{tabular}
\caption{\label{t:meanmedian} Mean and median of the dip-to peak rotation frequency change measurements (Hz)}
\end{table}

\end{document}